# Highly Itinerant Atomic Vacancies in Phosphorene

Yongqing Cai[†], Qingqing Ke[‡], Gang Zhang[†,*], Boris I. Yakobson[¶], and Yong-Wei Zhang[†,*]

[†]Institute of High Performance Computing, A*STAR, Singapore 138632
[‡]Institute of Materials Research and Engineering, A*STAR, Singapore 138634
[¶]Department of Materials Science and Nano Engineering, Rice University, Houston, Texas 77005, United States

*Supporting Information Placeholder*

**ABSTRACT:** Using detailed first-principles calculations, we investigate the hopping rate of vacancies in phosphorene, an emerging elemental 2D material besides graphene. Our work predicts that a direct observation of these mono-vacancies (MVs), showing a highly mobile and anisotropic motion, is possible only at low temperatures around 70 K or below where the thermal activity is greatly suppressed. At room temperature, the motion of a MV is sixteen orders faster than that in graphene, because of the low diffusion barrier of 0.3 eV. Built-in strain associated with the vacancies extends far along the zigzag direction while attenuating rapidly along the armchair direction. We reveal new features of the motion of di-vacancies (DVs) in phosphorene via multiple dissociation-recombination processes of vacancies owing to a small energy cost of ~ 1.05 eV for the splitting of a DV into two MVs. Furthermore, we find that uniaxial tensile strain along the zigzag direction can promote the motion of MVs, while the tensile strain along the armchair direction has the opposite effect. These itinerant features of vacancies, rooted in the unique puckering structure facilitating bond reorganization, enable phosphorene to be a bright new opportunity to broaden the knowledge of the evolution of vacancies, and a proper control of the exceedingly active and anisotropic movement of the vacancies should be critical for applications based on phosphorene.

## Introduction

Phosphorene, another elemental two-dimensional (2D) material in addition to graphene, is gaining increasing attention for potential applications in electronics and photonics.[1-5] It has been regarded as promising candidates of transistors owing to its remarkable mobility (~1,000 cm$^2$V$^{-1}$s$^{-1}$ at room temperature)[3] and tunable robust direct bandgap from 0.3 eV (bulk) to 1.5 eV (monolayer).[5] Similar to graphene, phosphorene can be synthesized by mechanical cleavage and liquid-phase exfoliation.[6,7] In spite of the fast growing knowledge of phosphorene,[8-11] the origin of its notoriously low stability still remains an important puzzle.[12,13] As one of the most stable polymorphs among all phosphorus allotropes,[14] according to the thermodynamic criteria, phosphorene starts to disintegrate quickly at ambient condition, which is generally ascribed to effects from external molecules.[15,16] While protecting phosphorene from the environmental influence via capping layers was found to effectively improve its stability,[17,18] intrinsic structural stability dictated solely by atomic defects like vacancies then becomes a fundamental descriptor for its applications. Recent compelling in-situ experiment in vacuum demonstrates a fast evolution of structure with temperature,[19] in which atomic sublimation occurring without melting process. However, it remains unclear what the roles of the vacancies play in the disintegrating process.

As one of the most important types of lattice imperfections, atomic vacancies govern various nanoscopic and mesoscopic processes in nanomaterials.[20-23] Topological defects in phosphorene tend to possess intriguing electronic properties and suitable for hosting foreign atoms.[24-28] Vacancies show a negative U behavior[29] absent in other well-studied semiconducting 2D materials. Normally, in 2D materials, their atomically-thin thickness is expected to facilitate the observation of these defects, reflected by the routine detections of vacancies in graphene[30-33] and MoS$_2$.[34,35] However, no observation of atomic vacancies in phosphorene sheet has been reported so far, despite several atomic-scale profile imaging efforts.[6,7,36] Moreover, how to control the motion of vacancies in phosphorene is still an open question.

Here by using first-principles calculations, we reveal an itinerant behavior of vacancies in phosphorene even at room temperature (RT). The motion of a mono-vacancy (MV) is sixteen orders faster than that in graphene, and its hopping rate is around three orders higher along the zigzag direction ($4.3 \times 10^7$ s$^{-1}$) than the armchair direction ($3.1 \times 10^4$ s$^{-1}$) at RT. We map out the right condition for imaging atomic vacancies in phosphorene: decreasing the measuring temperature down to 70 K or even lower. The exceedingly high motilities of MVs and di-vacancies (DVs) should account for the relatively low structural stability and high affinity to environmental molecules in phosphorene. We also demonstrate that moderate lattice strain can significantly modulate these activating barriers. This factor plays an important role in understanding the evolution of edge structure as strain can easily be built up at the edges due to the intrinsic edge stress.

## Computational Details

First-principles calculations within the framework of density functional theory (DFT) are performed by using VASP package.[37] We use the Perdew-Burke-Ernzerhof functional for the exchange-correlation potential and a kinetic energy cutoff of 500 eV. All the vacancies are modeled by using a 8×7×1 supercell (222 atoms for the perfect sheet) together with a vacuum layer with the thickness of 15 Å. The a= 3.335 Å and b= 4.571 Å are the relaxed lattice constants along the zigzag and armchair direction, respectively, for unit cell of phosphorene. The first Brillouin zone is sampled with a 3 × 3 × 1 Monkhorst-Pack grid. All the structures are fully relaxed until the forces exerted on each atom are smaller than 0.01 eV/Å. For the lattice subjected to the uniaxial strain, the relaxation of the lattice normal to the strained direction is considered. The formation energy ($E_f$) of a vacancy is calculated as $E_f = E_p - E_d - \sum_i n_i u_i$, where $E_p$ and $E_d$ represent the total energies of the perfect and defective system, respectively; $n_i$ is the number of the removed atoms of type $i$, and $u_i$ is the chemical potential of type $i$ atom. For phosphorene and graphene, $u_i$ is taken as the energy of a single atom in the perfect sheet. For MoS$_2$, $u_i$ is bounded by the formation of bulk Mo and gas phase of sulfur (eight-membered ring), corresponding to S-poor and S-rich conditions, respectively.

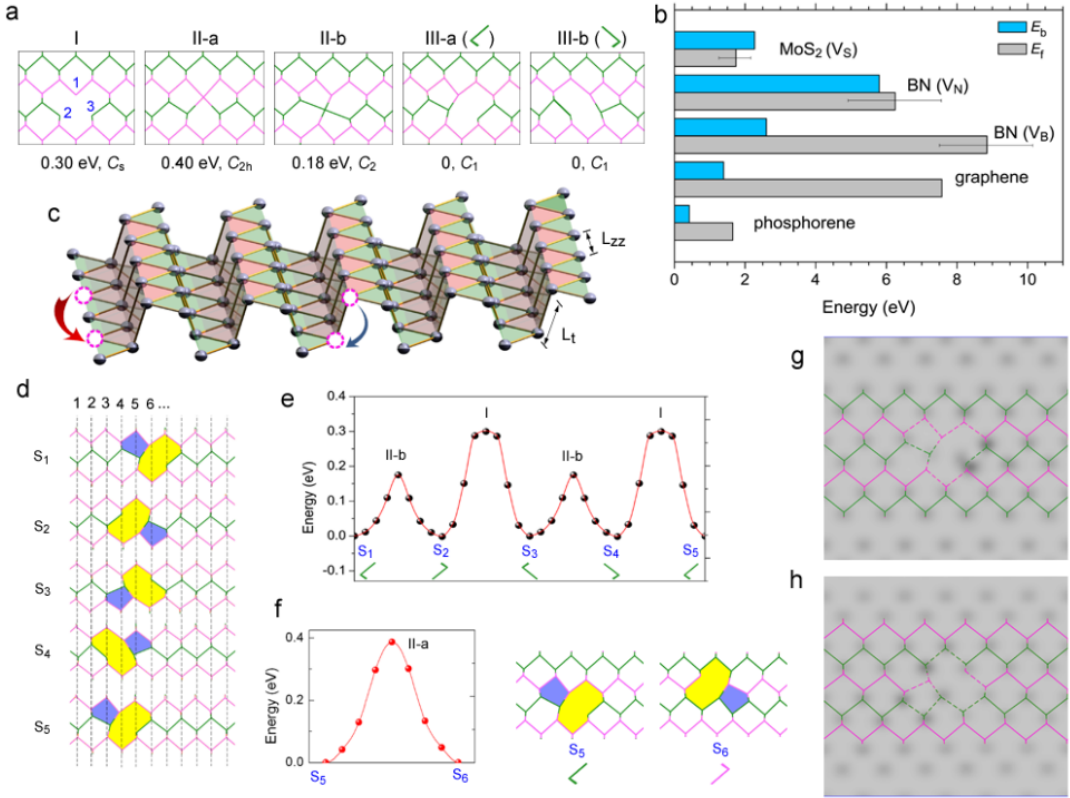

Figure 1. Energetics and kinetics of a MV in phosphorene. (a) Various states of MV: Type I, II-a and b structures are metastable, whereas the energetically degenerate type III-a and b structures, vividly represented by the rotated ticks, are the ground state. The local site symmetry and the $E_f$ of each configuration relative to the III-a and b are given. (b) Comparison of the $E_f$ and $E_b$ of a neutral MV between various 2D materials under dilute limit. For heteropolar layered materials like MoS$_2$ and BN, the $E_f$ of a particular type of vacancy is dependent on the chemical potential of the components, and this uncertainty is depicted by the horizontal line. The data of BN are adopted from Ref. [51] and [52]. (c) Atomic models for describing a MV (dashed circle) diffusing along the armchair (blue arrow) and zigzag (red arrow) directions. (d) Intermediates states (S$_1$-S$_5$) for a MV hopping along the zigzag direction between neighboring equivalent sites (from branch 6 to 4). (e) Energetic profiles obtained by NEB calculation for the migration of the vacancy from S$_1$ to S$_5$, where seven structural images are produced between two stable S$_i$ and S$_{i+1}$ (i=1, 2, .., 4) configurations. (f) Activation barrier for inter-chain vacancy diffusion from S$_5$ to S$_6$. Here the ticks are colored according to the site of the removed atom. (g) and (h) correspond to the STM images obtained at +1V of the bias voltage for structures with the MV located at the top ridge close to and the bottom ridge far from the tip, respectively.

## Results

A MV in phosphorene is formed by the removal of a single phosphorus atom, which creates three dangling atoms (see atoms 1, 2, and 3 in Fig. 1a) distributed in two neighboring ridges (Fig. 1a). Starting from this metastable MV (denoted as type-I), reconstructed vacancies can be formed by bridging the three peripheral atoms. Topologically equivalent metastable II-a and II-b structures are created by shifting atoms 1 and 2, respectively. Driven by Jahn-Teller distortion, the ground state of a MV occurs by pairing atom 1 with atom 2 or 3, leading to the formation of energetically degenerate lower symmetric III-a or III-b structures with pentagon-nonagon (59) pairs. Since the shape of the "9" ring in the type-III MV resembles a tick, in the following, rotated ticks such as "<" and ">" are used to denote various type-III MVs with different orientations and locations.

We compare the energetics of MV in phosphorene with those of sulfur vacancy (V$_S$) in monolayer MoS$_2$, boron vacancy (V$_B$) and nitrogen vacancy (V$_N$) in monolayer hexagonal BN (h-BN), and carbon vacancy (V$_C$) in graphene. The lattice constants of phosphorene, graphene, and MoS$_2$ used in our calculations are summarized in Table S2 in Supporting Information, which are consistent with those from literatures. The small difference (less than 2%) in the lattice constants between our work and the reference should be due to the different parameters adopted in DFT calculations. Our calculations show that the variation of the lattice constant around 2% is only able to change the magnitude of the migration barrier of mono-vacancy (shown below) by less than 0.05 eV. The values of $E_f$ are calculated under the same theoretical setting and compiled in Fig. 1b. The value of $E_f$ for MV in phosphorene is 1.65 eV, significantly smaller than that of graphene of 7.57 eV. Since the defect population depends exponentially on $E_f$, the equilibrium density of MV in phosphorene should be exceedingly larger than that of graphene under the same conditions. Amongst these 2D materials, only the value of $E_f$ for $V_S$ in monolayer $MoS_2$ (from 1.22 to 2.25 eV) is smaller than or comparable to that in phosphorene, while the values of $E_f$ for both $V_N$ (from 4.90 to 7.60 eV) and $V_B$ (from 7.50 to 10.20 eV) in h-BN are much higher than that in phosphorene. Here, the variability of the chemical potential (elemental rich and poor conditions) defines the upper and lower bounds of $E_f$. The much lower value of $E_f$ in phosphorene is associated with the intrinsically softer P-P bonds[27-29] compared with the much stronger C-C and B-N bonds and also the curvature effect related to the structural buckling.

Climbing image nudged-elastic band (NEB) calculations are performed to estimate the energy barrier ($E_b$) for the thermally activated migration of MV in phosphorene. Due to the unique puckered structure with parallel alignment of quasi-1D zigzag chains, diffusion of MV adopts two orthogonal paths (Fig. 1c): one is the intra-chain diffusion with MV hopping along the zigzag chain, and the other is the inter-chain diffusion with MV moving across the neighboring chains along the armchair direction. Combination of these two pathways permits the motion of MV in any direction in the sheet.

For the intra-chain diffusion of MV, the snapshots ($S_1$-$S_5$) and the corresponding energy landscapes are plotted in Fig. 1d and e. Three intermediate states ($S_2$, $S_3$ and $S_4$) are identified and the whole trajectory can be further decomposed into two basic steps (Fig. 1d and e): neighboring-site hopping ($S_1$-$S_2$ and $S_3$-$S_4$) with initial and end states (for instance, "〈" ⟷ "〉") transformed by a pseudo-inversion together with a shift of half a lattice ($L_{zz}/2$), and an on-site reorganization ($S_2$-$S_3$ and $S_4$-$S_5$) with transforming states (for instance, "〉" ⟷ "〈") connected with mirror symmetry. The values of $E_b$ are 0.18 and 0.30 eV and the transition states coincide with the II-b and I type MV states for the neighboring-site and on-site hopping processes, respectively. For the inter-chain diffusion (Fig. 1f), the MV climbs across the upper and bottom troughs with initial and final states ("〈" ⟷ "〉") connected with inversion, together with a glide of one tilted bond ($L_t$). The barrier is calculated to be 0.40 eV and the transition state adopts the II-a MV structure.

The sequence of $E_b$ for MV diffusion in the above 2D materials follows: phosphorene (0.40 eV) < graphene (1.39 eV) < $MoS_2$ ($V_S$, 2.27 eV) < BN ($V_B$, 2.60 eV) < BN ($V_N$, 5.80 eV). Therefore, phosphorene has the lowest $E_b$, which is only one third of that of graphene. According to the Arrhenius formula, the hopping rate ($v$) can be calculated by $v=v_s \exp(-E_b/kT)$, where $v_s$ is the characteristic frequency, normally around $10^{13}$ Hz, k is the Boltzmann constant and $T$ is the temperature. At RT, the rate-limiting process (inter-chain diffusion) of the migration of MV in phosphorene is estimated to be amazingly around 16 orders faster than that of graphene, signifying a significantly faster vacancy activity. By using the Vineyard formula, which allows the exact calculation of the prefactor $v_s$ (see Supplementary Table S3), the values of $v$ for MV in phosphorene at RT are obtained to be $2.5 \times 10^9$, $4.3 \times 10^7$, and $3.1 \times 10^4$ per second for the three basic motion modes, that is, the neighboring-site hopping along the zigzag direction, the on-site hopping along the zigzag direction, and the inter-chain diffusion along the armchair direction, respectively. Furthermore, we have conducted ab-initio molecular dynamics simulations to explore thermodynamic stability of MV (see details in Supporting Information) at finite temperature. For the MV at 300 K, several hopping events occur along the zigzag direction during a period of 7.3 ps (see Movie 1), consistent with the aforementioned hopping rate ($2.5 \times 10^9$) estimated based on the diffusion barrier.

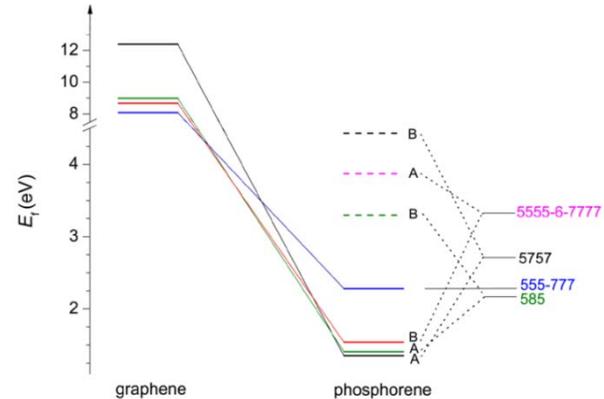

Figure 2. Energetics of di-vacancy in phosphorene. Opposite hierarchy of stability of DVs between graphene and phosphorene, and the lifting of structure degeneracy of DVs in phosphorene.

We show that these exceedingly high hopping rates even at RT imply a highly itinerant behavior of vacancies in phosphorene, which is absent in other 2D materials known so far. Moreover, motion of MV is found to be orientationally anisotropic: the value of $v$ along the zigzag direction is three orders higher than that along the armchair direction at RT. This anisotropy increases further with the reduction of temperature and the anisotropy ratio reaches up to seven orders at 100 K, implying the cross-ridge motion being significantly suppressed and all the hopping events merely occurring in the zigzag direction. Our prediction of the fast motion of vacancies along the zigzag direction may explain the recent experiment on the thermal decomposition of phosphorene under vacuum, showing the formation of "eye" shaped voids aligned along the armchair direction.[19]

According to Fig. 1d, the diffusion of MV along the zigzag direction involves continual rotation of 59 pairs together with shifts in the lattice. Owing to the same honeycomb topology, the above pattern of MV motion in phosphorene should also exist in graphene. Indeed, MV diffusion with rotation plus lattice shifts along the zigzag direction was previously captured by TEM in graphene.[38] Unfortunately, due to the nearly angularly isotropic diffusion of MV, this phenomena in graphene was overlooked. Our finding in phosphorene provides cogent evidence for the presence of this intriguing mechanism and elucidates its role in structural evolution in these 2D materials. In principle, the significantly anisotropic motion and the puckered structure enabling sharp protrusions in STM images of MV (Fig. 1g and h) could facilitate the observation of quick succession of MV in phosphorene.

Due to the highly itinerant behavior of MVs, frequent coalescence of two isolated MVs into a DV can occur even at moderate temperatures in phosphorene. For an ideal honeycomb lattice like graphene, a fully compensated DV structure has several forms, including the pentagon-octagon-pentagon (585), triple pentagon-triple heptagon (555-777), and quadruple pentagon-hexagon-quadruple heptagon (5555-6-7777) types. Owing to the structural anisotropy, more types of DV structures could appear in phosphorene to accommodate the two missing atoms. Indeed, we find that each predecessor of a DV in graphene without a threefold ($C_3$) symmetry, such as 585 and 5555-6-7777 type defect, evolves into two topological equivalent structures differing in the relative orientation of the defect cores with the ridges in phosphorene (see Fig. S2). To distinguish these two states, in the following, a DV with its characteristic geometric profile (such as pentagon-nonhexagon group) lying along the diagonal direction is denoted as "A" type while its sister conformation along the orthogonal (armchair and zigzag) direction, generally with a higher symmetry, is termed as "B" type. Such lifting of structural degeneracy does not occur in the 555-777 DV due to the presence of the pseudo $C_3$ symmetry.

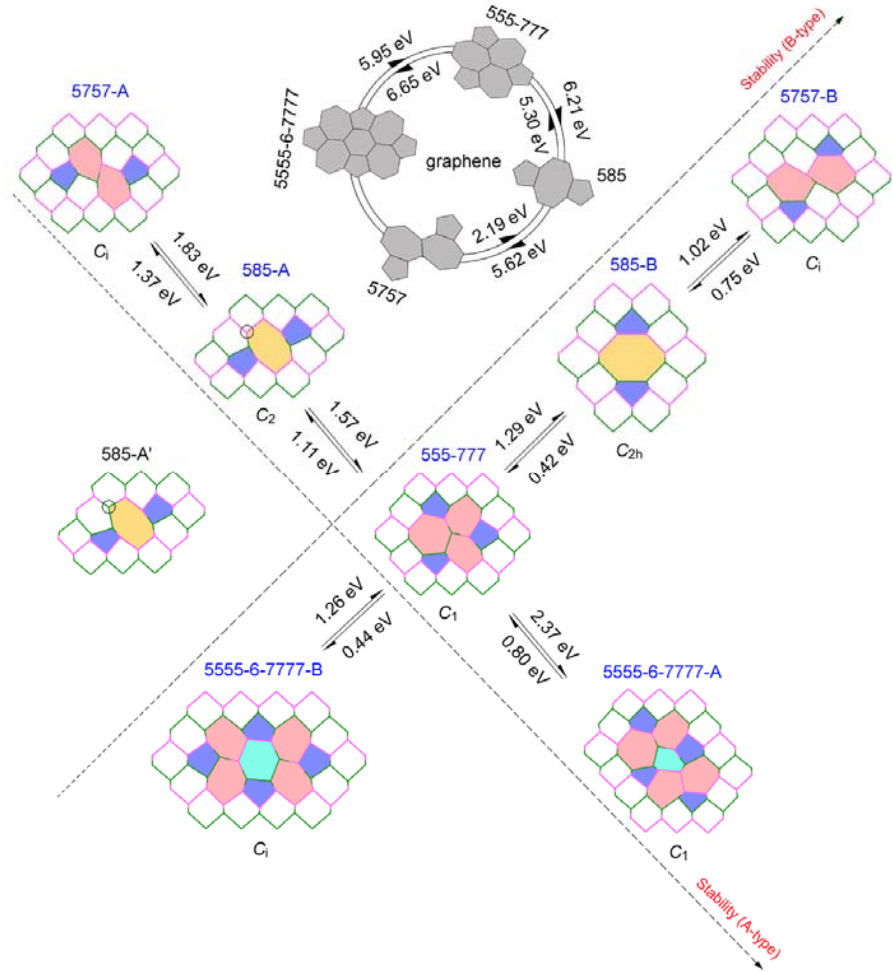

Figure 3. Transformation between different DV structures. Diagram shows the diffusion barriers (given by the numbers above the arrows) for forward and backward transformations between various DVs. The stability for the A-type and B-type DVs decreases in the order along the dashed arrows. The local site symmetry is given below the atomic model with pentagons colored in blue, hexagons in cyan, heptagons in violet and octagons in yellow. The barriers for transformations between various DVs in graphene are also plotted in the top inset for comparison. The left inset presents the 585-A' DV, which is obtained through transforming the circled atom in 585-A from the lower ridge (pink color) to the opposing top ridge (green color).

For graphene, it is well established that the 555-777 DV is the most energetically favored, followed by the 5555-6-7777 and 585 DVs.[39] The sequence of their $E_f$ values follows 555-777 (8.09 eV) < 5555-6-7777 (8.69 eV) < 585 (8.99 eV) < 5757 (12.4 eV). Surprisingly, for phosphorene, the ordering of the stability of the above topological defects is completely reversed with 555-777 (2.28 eV) > 5555-6-7777-B (1.54 eV) > 585-A (1.41 eV) > 5757-A (1.35 eV) and much smaller $E_f$ compared with graphene. The 5757 structure, rarely observed in graphene owing to its large $E_f$, is the lowest energy configuration in phosphorene. The $E_f$ of the "A" and "B" configurations can differ up to 3.08, 1.89, and 2.33 eV for 5757, 585, and 5555-6-7777, respectively (Fig. 2a), largely due to the different strain relief between A- and B-typed DVs and electron-lattice coupling. Interestingly, our ab initio molecular dynamics simulations show that the 5757-A vacancy structure is very stable, and there is no change in its topology in a period of 10 ps of MD calculation at 800 K. (see Movie 2 in Supporting Information)

We note that a DV in graphene has an obviously larger value of $E_f$ (8.09 - 12.4 eV) than that of MV (7.57 eV). In contrast, a DV in phosphorene has a comparable or even smaller value of $E_f$ (1.35 - 2.28 eV) than a MV (1.65 eV), explaining the sublimation observed in phosphorene at elevated temperatures.[19] Thus, two adjacent MVs are energetically favored to coalesce to form a DV. The release of energy for merging two isolated MVs to form a DV (5757-A), or equivalently, the energy cost to dissociate a DV into two isolated MVs, is estimated to be 1.05 eV, which is obtained by calculating the interaction energy $E_i = E_{DV} - 2E_{MV} + E_p$, where $E_{DV}$, $E_{MV}$, and $E_p$ are total energies of a supercell containing a DV, a MV, and without any vacancy, respectively. All the above features in phosphorene are rooted in its sp$^3$ bonding character, which can accommodate large variations of bond length and bond angle caused by the defect-associated lattice deformation.

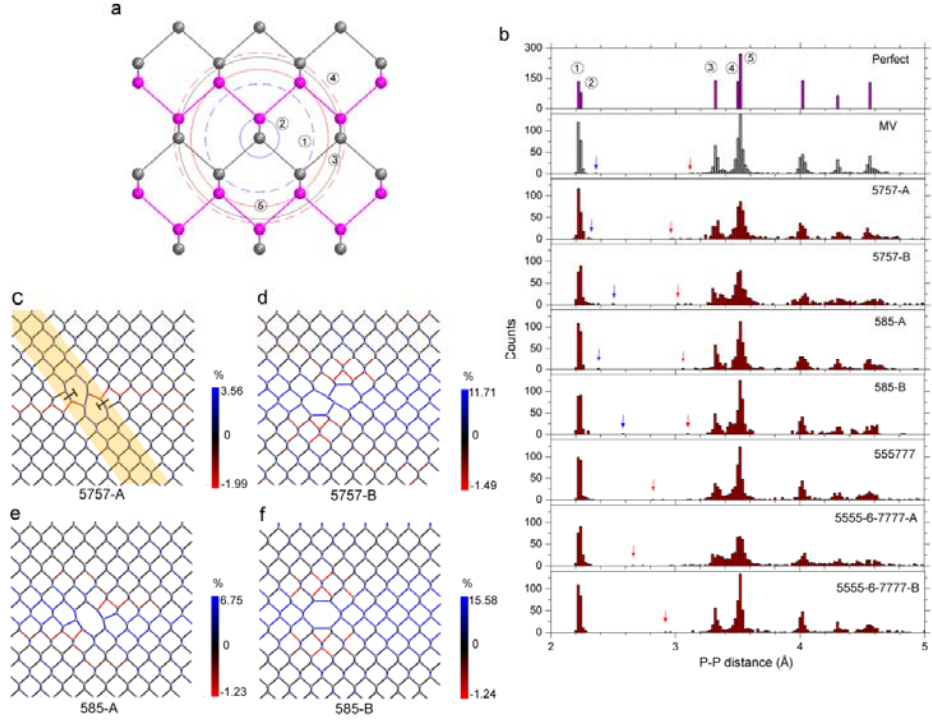

Figure 4. Structural characteristics and strain field associated with various vacancies. (a) Local atomic environment for a phosphorus atom. Numbered circles, according to the P-P distance given in (b), indicate the surrounding atoms from the nearest to 5$^{th}$ nearest neighbors. (b) Histogram of P-P distances for perfect and defective phosphorene. The blue (red) arrows indicate the occurrence of very large (short) P-P distances for nearest (the second nearest) neighbors at the defect core. (c)-(f) Strain fields around the 5757-A, 5757-B, 585-A, and 585-B DV cores. Bonds are colored according to the variation of the bond length to visualize the strain field around the vacancy center.

The energy barrier $E_b$ for the transformations between different DVs is plotted in Fig. 3. For the purpose of comparison, the forward and backward energy barriers between different DVs in graphene (5555-6-7777, 555-777, 585 and 5757) were also calculated (the inset of Fig. 3 and refer to details in Supplementary Fig. S3). In phosphorene, owing to its buckled honeycomb structure, the circular pathway shown in the inset of Fig. 3 in graphene is split into two paths corresponding to the "A" and "B" typed DVs, respectively, and the 555-777 acts as the connecting node.

It is well-known that the migration of DV in the graphene can be achieved through a series of Stone-Wales-type transformations among various DVs.[39] The $E_b$ for the transformations between the low energy DVs (585, 555-777 and 5556-6-7777) are significant (greater than 5 eV). The large barriers suggest that DVs in graphene are locked in a specific position and orientation at RT once formed, and the thermally activated motion of DV requires an extremely high temperature of 3000 K.[40,41] In contrast, our results show that transitions of DVs in phosphorene have much lower $E_b$ ranging from 0.44 to 1.83 eV (estimated hopping rate from $2.0 \times 10^9$ to $6.8 \times 10^{-3}$ per second at 600 K) for the four low energy DVs (5757-A, 585-A, 555-777 and 5555-6-7777-A), implying a much enhanced thermal activity even at RT. Interestingly, these $E_b$ for DV transitions are comparable to those of MV motion. This is in strong contrast to other 2D materials known so far, which show a frozen character and an extremely low mobility of DVs. Our calculations reveal that rotation and migration of DVs in phosphorene can occur frequently around 600 K. These low barriers can be ascribed to the small energy costs for bond rotation and torsion. From the energetic point of view, motion of a large vacancy cluster in phosphorene can occur in a new way, that is, through repeatedly attaching and detaching the more rapidly moving MVs. This is due to the small energy cost for the MV detachment from the vacancy clusters, for instance ~ 1.05 eV for the separation of two MVs from a DV. Such process in graphene is impossible due to the high energy cost for MV detachment (~ 7 eV) from a vacancy cluster.

Based on our calculations, we identify the following features of DV migration in phosphorene. First, the small $E_f$ of 5757 signifies its large population. In graphene, although the 5757 DV plays an important role as the intermediate state in the migration of DVs, it can only be produced by electronic irradiation owing to its extremely large $E_f$ (12.4 eV).[39] Second, the local buckling of the honeycomb structure allows easy bond flip and concurrently more intermediate states. For instance, for the transitions from 585-A to 5757-A or 555-777, the 585-A first evolves into a low-symmetric structure (585-A' structure, see the left inset of Fig. 3) by shifting one of the peripheral atoms at the octagon to the opposing ridge. This lowers the barrier of the transition from 585-A to 5757-A (555-777) by 0.96 eV (1.57 eV) compared with that of a direct rotation of the "shoulder" (the side) bond of the octagon (see supplementary Fig. S4). Similar intermediate states are also found during the transitions between 555-777 and 585-B, and between 585-B and 5757-B.

Finally, we analyze the modification of structures induced by these topological defects. Fig. 4a shows the local atomic environment for a P atom. The histograms of P-P distances in perfect and defective phosphorene containing MV, and the two lowest energy DVs (5757-A and 585-A), and their topologically equivalent conformers (5757-B and 585-B) are plotted in Fig. 4b. For the perfect lattice, the left most peak, which consists of a couplet with two types of bonds, is related to the nearest neighboring bonding along the zigzag (2.22 Å) and armchair (2.24 Å) directions, respectively. Upon the introduction of vacancies, a broadening of these lines occurs, indicating a distortion of the lattice. Notably, there exist several isolated lines corresponding to elongated P-P bonds indicated by the blue arrows in Fig. 4b. These lines correspond to the newly reconstructed bonds shared by the hexagon and heptagon in the 5757-A and -B DVs, or the pentagon and nonhexagon in the MV and the 585-A and -B DVs. Broadening also occurs in the second nearest and other far neighbors. The occurrence of the new short P-P distances (marked by red arrows) around 3 Å indicates a significant flattening of the puckered lattice in the proximity of the defect core, thus modifying the inter-ridge coupling.

Figure 4c-f illustrates the strain field around the core of DVs. Significant local tensile strains up to 3.56% (11.71%), and 6.75%

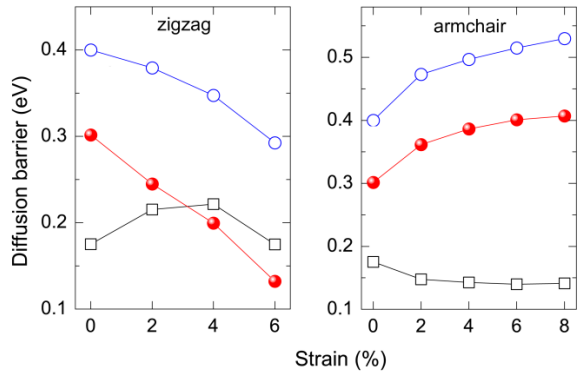

Figure 5. Modulation of the barriers for the neighboring-site (square), on-site (filled circle), and inter-chain (open circle) hoppings of MV by applying uniaxial strains along zigzag and armchair directions.

(15.58%) are observed for the 5757-A(-B) and 585-A(-B), respectively. Compressive strains are distributed around the head of pentagons. Generally, the B-type DVs present a larger lattice deformation than their A-type counterparts, consistent with the distance histogram shown in Fig. 4b. Also, the strain field is highly anisotropic with the strain extending far along the zigzag direction while attenuating rapidly merely around one unit cell along the armchair direction.

It is expected that these large lattice distortions may exert marked effects on the energetics and kinetics of other defects nearby. Indeed, our calculations show that the strain can greatly affect the diffusion barrier of MVs (Fig. 5). Overall, uniaxial tensile strains along the zigzag direction promote the motion of MVs (A 2% strain leads to a 17% decrease in the on-site hopping barrier), while the tensile strain along the armchair direction has an opposite effect by suppressing the motion of MV, thus enabling the tuning of MV dynamics via strain engineering. In addition, it is known that the motion of pentagon-heptagon (57) dislocation dipoles accounts for the superplasic behavior in $sp^2$ carbon materials at high temperature.[42] The significantly smaller energy cost for losing an atom and the much lower energy barrier for the bond flipping in phosphorene should greatly facilitate the gliding and climbing of 57 dislocations along the diagonal direction (Fig. 4c). Hence, it is expected that the superplastic deformation occurring in graphene at high temperatures should occur in phosphorene in moderate temperature, and a distinctively different thermal-structural-mechanical relationship may exist.

**Discussion**

We have shown that the neighboring-site hopping occurs around every 400 ps at RT. Since scanning tunneling microscopy (STM) and scanning electron microscopy (TEM) require much longer imaging timescale, generally, on the order of second, direct observation of MVs at RT is impossible due to the dispersive image arising from the superposition of different reconstructed MV states. Our calculations show that by reducing the temperature to liquid nitrogen (around 70 K), it is possible to elongate the hopping time up to 2 seconds, allowing for clear STM or TEM imaging. Our simulated STM images for MV and DV are shown in Fig.1g and h and Fig. S2, respectively. In reality, residual chemical species introduced during fabrication or applications,[7,9,43,44] may impede the movements of phosphorus atoms and induce slower hopping or even locked configurations of vacancies as demonstrated in hydrogen passivated MV in graphene.[33] In phosphorene, we have investigated the effect of an important chemical species-oxygen on the kinetic behavior of MV by using ab initio molecular dynamics (MD) calculations at 300 K. (see Movies in Supporting Information) We consider two most likely adsorbing states of oxygen: physisorbed $O_2$ molecule and chemisorbed O atom in the MV center. Our results show that the physisorbed $O_2$ molecule (initially ~ 3 Å above the MV) interacts only weakly with MV and cannot be trapped by the MV (see Movie 3). For the chemisorbed O atom, it is initially bonded to passivate the unsaturated P atom in the MV. Surprisingly, the MD results show that the chemical adsorption of O only has minor effect on the mobile behavior of MV since several hopping events of MV are still observed (see Movie 4). Therefore, the itinerant behavior of MV is retained in oxygen environment. The itinerant character of vacancies in phosphorene can be ascribed to two reasons: the first is the presence of the lone-pair electrons in the phosphorus atoms, which favors the reorganization of bonds during vacancy hopping; the second is related to its puckered structure, which facilitates the bond rotation and reduces the energy cost for local structural bending, compression or tension.

We note that previous studies ascribed the low stability of phosphorene to the adsorption of environmental molecules on perfect phosphorene surfaces.[12,13,15] Here our study suggests that the low chemical stability of phosphorene may be rooted in the intrinsic itinerant behavior of atomic vacancies. In addition, the intrinsic stability may also be affected by the itinerant vacancies. A recent in-situ experimental study observed "eye" shaped voids in phosphorene in vacuum at around 700 K.[19] These "eye" shaped voids are likely the consequence of the coalescence of the exceedingly mobile atomic vacancies through the strongly direction-dependent diffusion. Since there will be edge stress at the edges or dislocations which is absent the perfect interior region, any degradations of phosphorene are likely to initiate at the edges due to a smaller barrier with strain shown in Fig. 5. Therefore, we suggest that structural degradation of phosphorene may initiate at edges and thus proper passivation and protection of the edges may be effective to enhance the stability.

For any attempts to grow phosphorene via chemical vapor deposition (CVD),[45] the very active characteristics of the vacancies imply the need for a careful control of atomic chemical potential. Since CVD growth is typically far away from thermodynamic equilibrium, a proper selection of temperature is required in order to control atomic kinetic behavior. For example, the growth temperature should be high enough to overcome the energy barrier for P atoms to attach to the growth front, meanwhile it should be low enough to suppress the disordering arising from the coalescence of mobile vacancies.

On the other hand, the ultrafast mobility of atomic vacancies in phosphorene may potentially lead to new technological possibilities. For instance, mobile vacancies accumulated in regions of filament or grain boundaries in memresistive semiconductors are the underlying origin for resistive switching devices.[46-49] However, fabrication based on either conventional oxides[46,47] or recently developed anion-deficient 2D layered transition metal dichalcogenides[48-50] requires a properly aligned conducting filaments or grain boundaries connecting the electrodes, which clearly complicates the fabrication of those devices. Moreover, as shown in resistive switching devices based on oxide bulk materials and 2D MoS$_2$,[48-50] aggregative vacancies are critically important for the formation of filaments, which are highly desired for high-performance resistive switching devices. Therefore, the highly mobile and strongly directional moving vacancies in phosphorene are promising for developing such novel devices.

**Summary**

In conclusion, remarkably high translational and rotational mobilities of atomic vacancies in phosphorene are demonstrated. These exceedingly high motilities of MVs explain the difficulty in imaging them at ambient conditions. Meanwhile, the itinerant vacancies can explain the fast structure evolution of phosphorene in vacuum and may be responsible for the high chemical affinity of phosphorene to environmental molecules at elevated temperatures. Resembling its predecessor, graphene, which is renowned for its ultrahigh mobile electrons/holes, phosphorene possessing the highly mobile vacancies offers new possibilities in defect engineering and manipulation without the need of continual electron irradiation as required in other 2D materials known so far. Potential control, condensation and ordering of these itinerant vacancies under external stimuli such as thermal or electric field should be highly important for applications.

## ASSOCIATED CONTENT

**Supporting Information**

NEB results of diffusion of MV in graphene and MoS$_2$; prefactor for the diffusion and hopping rate of MV of phosphorene at various temperatures; diffusion pathway of transformations between various DVs in graphene and phosphorene; The movies for describing the structural evolution of mono-vacancy at 300 K (Movie 1) and that of di-vacancy at 800 K (Movie 2); The movies for showing the effect of oxygen molecule (Movie 3) and atom (Movie 4) on the kinetic behavior of the MV at 300 K. This material is available free of charge via the Internet at **http://pubs.acs.org**.

## AUTHOR INFORMATION

**Corresponding Author**

zhangg@ihpc.a-star.edu.sg; zhangyw@ihpc.a-star.edu.sg

**Notes**
**The authors declare no competing financial interests.**

## ACKNOWLEDGMENT

This work was supported in part by a grant from the Science and Engineering Research Council (152-70-00017). The authors gratefully acknowledge the financial support from the Agency for Science, Technology and Research (A*STAR), Singapore and the use of computing resources at the A*STAR Computational Resource Centre, Singapore.